\documentclass[twocolumn, 5p, 12pt]{elsarticle}

\usepackage{graphicx}
\usepackage{subfigure}
\usepackage[dvipdfx,cmyk]{xcolor}

\usepackage{amssymb}

\biboptions{sort&compress}

\journal{Physics Letters A}

\begin{document}

\begin{frontmatter}



\title{PageRank model of opinion formation on Ulam networks}

\author[yerphi,dijon,ashtarak]{L.~Chakhmakhchyan}
\ead{levonc@rambler.ru}
\ead{tel: +37493331861}
\ead{fax: +37410344736}

\author[dima]{D.~Shepelyansky}
\ead{dima@irsamc.ups-tlse.fr}
\ead[url]{http://www.quantware.ups-tlse.fr/dima}

\address[yerphi]{A.I.~Alikhanyan National Science Laboratory, 0036 Yerevan, Armenia}

\address[dijon]{Laboratoire Interdisciplinaire Carnot de Bourgogne, UMR CNRS 6303, \\ Universit\'{e} de Bourgogne, 21078 Dijon Cedex, France}

\address[ashtarak]{Institute for Physical Research, 0203 Ashtarak-2, Armenia}

\address[dima]{Laboratoire de Physique Th\'{e}orique du CNRS, IRSAMC, Universit\'{e} de Toulouse, UPS, F-31062 Toulouse, France}

\begin{abstract}
We consider a PageRank model of opinion formation on Ulam networks,
generated by the intermittency map and the typical Chirikov map.
The Ulam networks generated by these maps have  certain similarities with
such scale-free networks as the World Wide Web (WWW),
showing an algebraic decay of the PageRank probability.
We find that the opinion formation process on Ulam networks
have certain similarities but also distinct features
comparing to the WWW. We attribute these distinctions to
internal differences in network structure of the Ulam and WWW networks.
We also analyze the process of opinion formation in the frame of
generalized Sznajd model which protects opinion of small communities.
\end{abstract}


\end{frontmatter}

\section{Introduction}
\label{intr}

The understanding of mechanisms of opinion formation in the modern society is at
the heart of a newly emerged research field, known as sociophysics \cite{socio}.
A number of voter models has been developed during the last few decades
for understanding of nontrivial features of opinion formation
in a society (see Refs.~\cite{model, model1, model2, model3, model4} for details).
However, these models are generally considered on abstract regular lattices,
which are very different from a scale-free structure of modern social networks
with hundreds of millions of users.
In particular, such social networks as LiveJournal \cite{LJ},
Facebook \cite{FB} or Twitter \cite{TW} allow to have a rapid information exchange
over a large fraction of network users and to share social events,
making an essential contribution to the mass opinion formation.
These social networks have a growing influence on the social and political life.

A straightforward way of taking into account the main features of such
networks was recently proposed in Ref.~\cite{prof}:
the opinion on each given node of a scale-free network is assumed to be formed
by opinions of its linked neighbors, weighted with their PageRank probability.
The latter quantity is interpreted as a probability of finding a random surfer on
a given node \cite{google, google1}. Obviously, this approach introduces the notion of
importance of a node, naturally reproducing the real society, where each person
has its degree of authority. Mathematically the PageRank is defined as
the right eigenvector with unit eigenvalue of Google matrix of a given network
\cite{google1}.
Although the PageRank algorithm was initially proposed for an efficient
ranking of web pages \cite{google}, it turned out to be useful for
the analysis of broad class of real
networks including e.g. scientific journal rating, neuronal and world trade networks,
etc. \cite{journal0, journal, neuron, trade}.
The rules of Google matrix construction for a given directed network
are described in \cite{google, google1, neuron}.

In the present work we study the PageRank Opinion Formation (PROF) model,
proposed in \cite{prof},
on another family of directed networks, known as Ulam networks.
The Ulam method, introduced in Ref.~\cite{ulam},
was initially proposed for constructing a matrix approximant
for a Perron-Frobenius operator of dynamical systems
(we note that the Google matrix also falls in the same class of operators).
The Ulam conjecture \cite{ulam}
was shown to be true for various types of generic fully chaotic maps on
an interval \cite{maps, maps1, maps2, maps3}.
Recent studies have shown that this method naturally generates a class of directed networks,
which properties have certain similarities with the WWW directed networks
\cite{ulamd, ulamd1}. Thus the Ulam networks demonstrate
a sensitivity to the damping parameter
$\alpha$ of the corresponding Google matrix and a power law decay of its PageRank.
Here we are interested in two particular examples: the typical Chirikov map with dissipation
and the one dimensional intermittency map. The first one, introduced in Ref.~\cite{chir}
for a description of continuous chaotic flows,
has been studied in \cite{frahm,ulamd}.
The second one is generated from intermittency maps,
studied in systems exhibiting intermittency phenomenon,
featuring anomalous diffusion and transport \cite{pomeau,junction, artuso, artuso1, pikovsky}.  We note that a similar approach, directly related to the Ulam method and based on network representation of coarse-grained maps, can be used for the investigation of predictability and information aspects of a system \cite{comm}. In addition, coarse graining and associated symbolic dynamics, local and global spectra analysis is also at the heart of prediction, error estimates and monitoring of nonlinear complex systems \cite{bif, nicoli}.

In this work we analyze the properties of PROF model on the Ulam networks
and study the influence of network elite on opinion formation process.
We also consider the Sznajd model \cite{sznajd}, generalized for scale-free networks
following \cite{prof}. This model incorporates the effect of groups,
consisting of voters of the same opinion
following the trade union slogan
{\it united we stand, divided we fall}. We note that the above models share some similarities with a recently analyzed model of continuous opinion dynamics \cite{contineous}. In particular, the diffusion effects introduced there can be associated with the damping factor of the Google matrix, to be discussed below.

In the rest, the paper is organized as follows:
in the next section we give a brief description of the Ulam method
and PROF model and present our numerical results.
In Section \ref{sznajd} we combine the PROF and Sznajd models and
analyze their properties on Ulam networks. The discussion of the results is
given in Section \ref{concl}.

\section{The PROF model and Ulam networks}
\label{prof}

We start with a brief outline of the Ulam method for dynamical maps
following the description given in \cite{ulamd, ulamd1}.
As the first model we use the one-dimensional (1d) intermittency map
described in \cite{ulamd1}:
\begin{equation}
{\bar x}=f(x)=\left
\{\begin{array}{c c }
x + (2 x)^{z_1}/2, \; \mathrm{for} \; 0\leq x<1/2  \\ \nonumber
(2 x - 1 - (1 - x)^{z_2} + 1/2^{z_2})/ \\ (1 + 1/2^{z_2}), \; \mathrm{for} \; 1/2 \leq x<1
\end{array} \right.\label{1}
\end{equation}
where ${\bar x}$ notes the new value of variable $x$.
The Ulam network generated by this map is constructed in the following way:
the whole interval $0<x<1$ is divided to $N$ equal cells and $N_c$ trajectories
(randomly distributed inside a cell) are iterated on one map iteration from cell $j$,
to obtain matrix elements for transitions to cell $i$: $S_{ij}=N_i(j)/N_c$,
where $N_i(j)$ is the number of trajectories arrived from cell $j$ to cell $i$.
From the matrix $S_{ij}$, one constructs the Google matrix \textbf{G}, defined as:
\begin{equation}
{\bf G}=\alpha {\bf S} + (1-\alpha){\bf E}/N, \label{G}
\end{equation}
where $E_{ij}=1$ and $\alpha$ is the damping factor. We use a probability normalization
of the eigenstate $|\psi_1\rangle$ (with a unit eigenvalue) of the matrix (\ref{G}),
which results in the PageRank $P_j$
of the network  (see \cite{ulamd1} for a detailed description of its properties).
We also arrange all $N$ nodes in monotonic decreasing order of the PageRank probability.
In what follows we set the damping factor of
the Google matrix of the intermittency map (\ref{1}) to $\alpha=1$.
We also fix the parameters of (\ref{1})
to $z_1=2$ and $z_2=0.2$. This choice gives a power law decay
of the PageRank (sorted in descending order): $P_j\propto1/j$ \cite{ulamd1}.

We construct  the PROF model for the Google matrix of the intermittency map (\ref{1})
in the following way. We associate each node of the network with a spin variable $\sigma_i$,
taking values $+1$ (red color) or $-1$ (blue color).
Afterwards, we compute the quantity $\Sigma_i$ over
all directly linked neighbors $j$ of a node $i$:
\begin{eqnarray}
\Sigma_i=a\sum_jP_{j, in}^++b\sum_jP_{j, out}^+ \\ \nonumber
-a\sum_jP_{j, in}^--b\sum_jP_{j, out}^-,
\end{eqnarray}
where $P_{j,in}$ and $P_{j,out}$ denote the PageRank probability $P_j$ of a node $j$
pointing to node $i$ (incoming link) and a
node $j$ to which node $i$ points to (outgoing link). The two parameters
$a$ and $b$ are used to tune the importance of incoming and outgoing links
with the imposed relation $a+b = 1$ ($0<a, b<1$).
The values $P^+$ and $P^-$ correspond to red and blue nodes respectively.
On one iteration the value of a spin $\sigma_i$ is fixed to $+1$ (red)
for $\Sigma_i>0$
or $-1$ (blue) for $\Sigma_i < 0$.
We note that the $a$ and $b$ parameters define the type of a society:
for a large value $a$ a person takes mainly the opinion of those electors
who point to him/her (a tenacious society)
and the opposite for large values of $b$ (a conformist society).

\begin{figure}[h]
\begin{center}
\includegraphics[width=6cm]{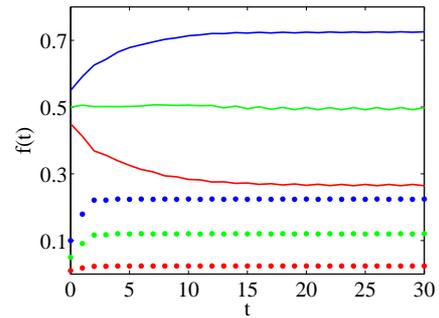}
\caption {Time evolution of the opinion, given by a fraction of red nodes $f(t)$,
as a function of number of time iteration $t$ ($a=b=0.5$). Full curves correspond
to different initial fractions $f_i=f(0)$ at a random realization:
$f_i=0.45$ (red); 0.5 (green); 0.55 (blue). The dotted curves stand for
the initial state with the first $N_{top}$ nodes of the highest PageRank probability
being red: $N_{top}=100$ (red); $N_{top}=500$ (green); $N_{top}=1000$ (blue).
The total matrix size is $N=10^4$; $\alpha=1$.
\label{fig1}}
\end{center}
\end{figure}

In Fig.~\ref{fig1} we present the evolution of
the fraction of red nodes $f(t)$ ($f(t)=N_{red}/N$)
versus the iteration time $t$. We distinguish two important cases, namely,
when initially opinions are randomly distributed over the network, and
when the first $N_{top}$ nodes of the highest PageRank
probability are of the same opinion,
e.g. of a red color. For a random distribution the system converges to
its final state after $t_c\approx 25$ iterations for  $a=b=0.5$.
Iterations are defined as in \cite{prof}.

In Fig.~\ref{fig1} we show the time evolution of opinion for the initial state
where the society elite, corresponding to the top nodes $N_{top}$
of highest PageRank probability, has
the same opinion (dotted curves). In this case the elite can impose its opinion to a
faction of society which is by a factor $2-3$ larger than the initial fraction.
However, in comparison with the social or university networks
considered in \cite{prof} this increase is less significant that is due to a smaller number
of linked nodes for the Ulam network of intermittency map.

\begin{figure}[ht]
\begin{center}
\includegraphics[width=0.23\textwidth]{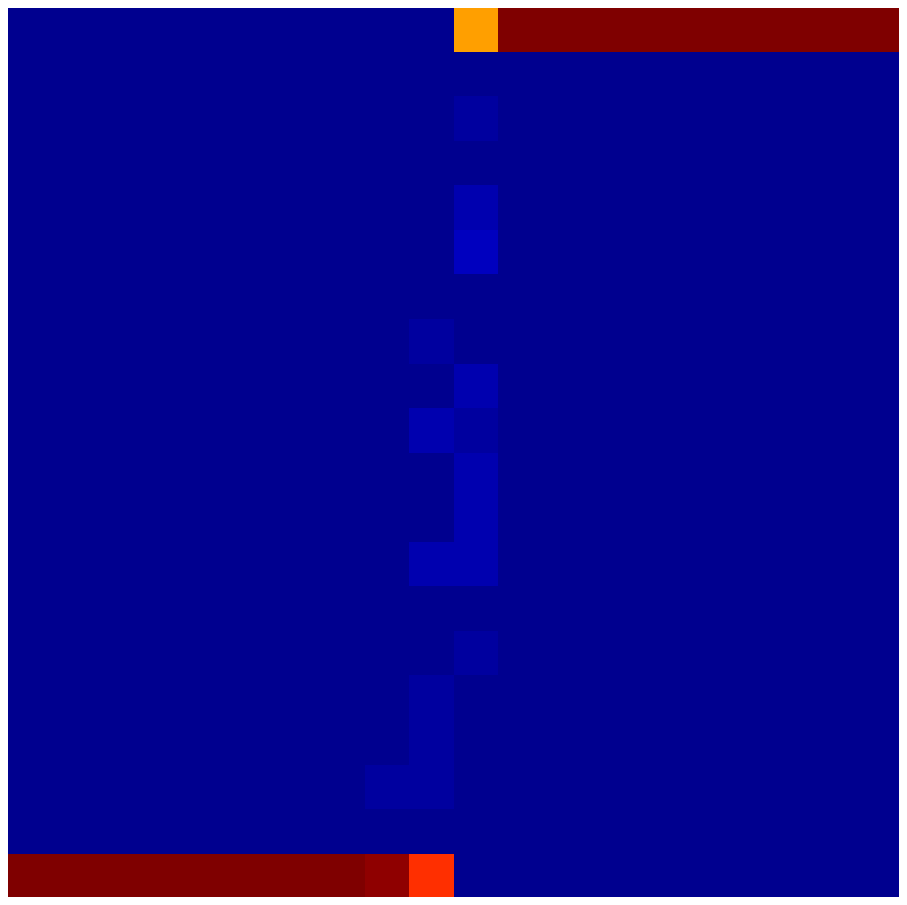}
\includegraphics[width=0.23\textwidth]{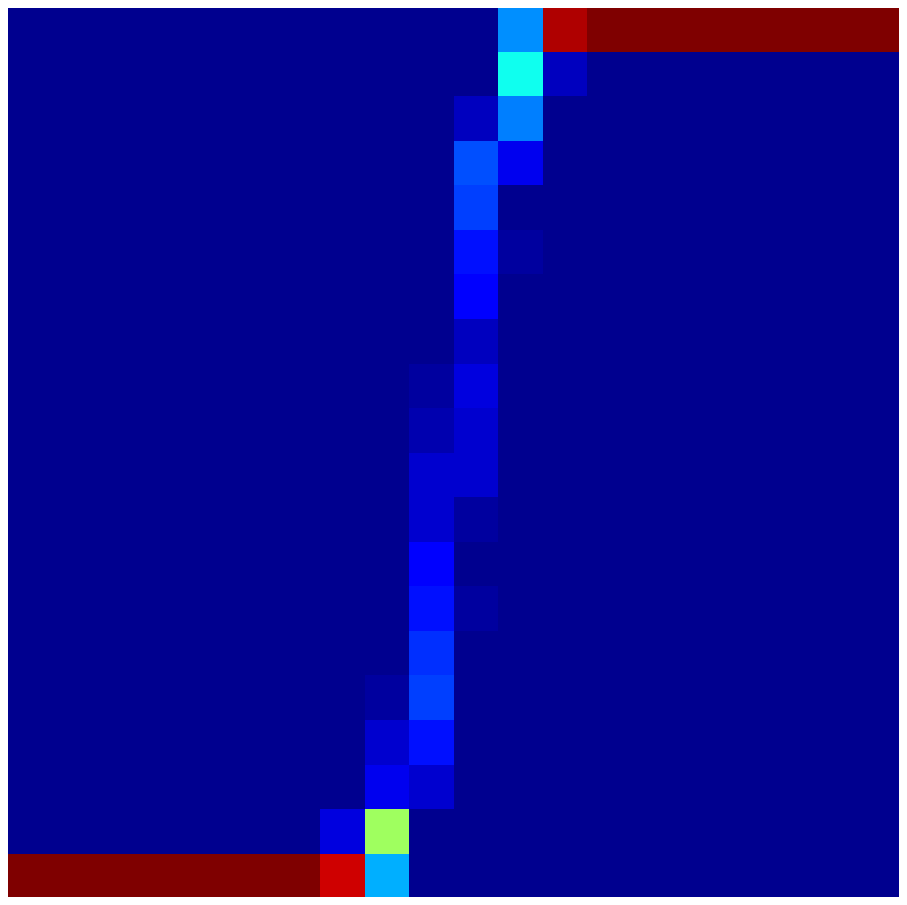}\\
\includegraphics[width=0.23\textwidth]{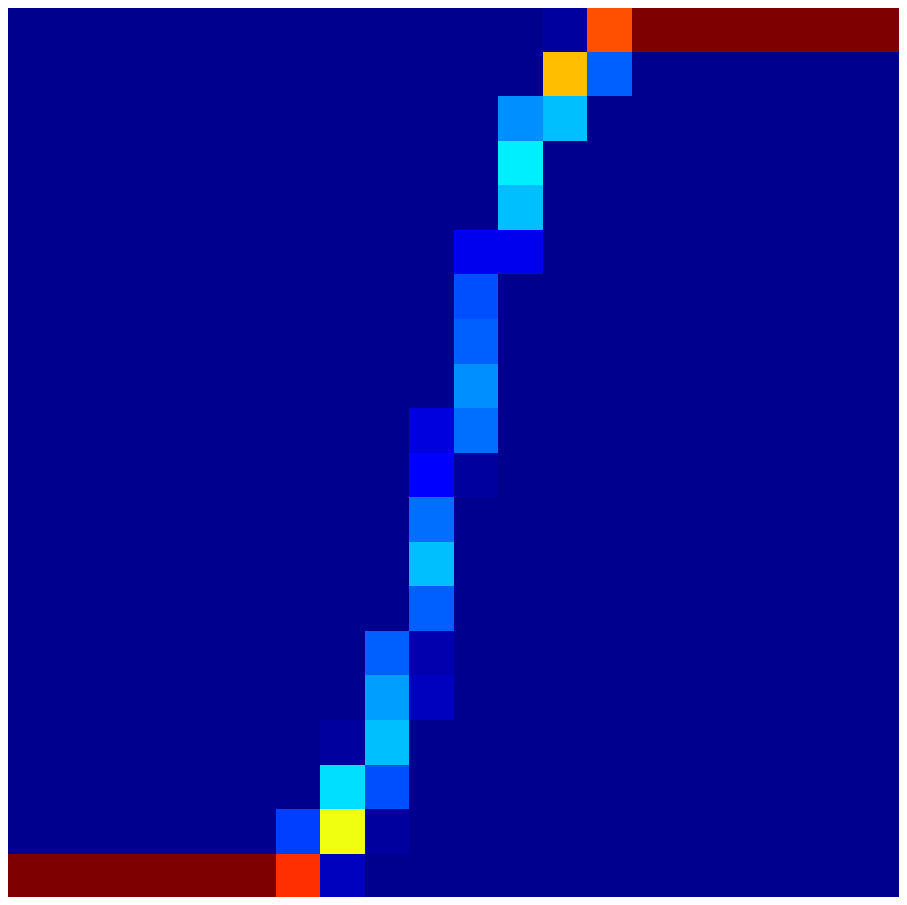}
\includegraphics[width=0.23\textwidth]{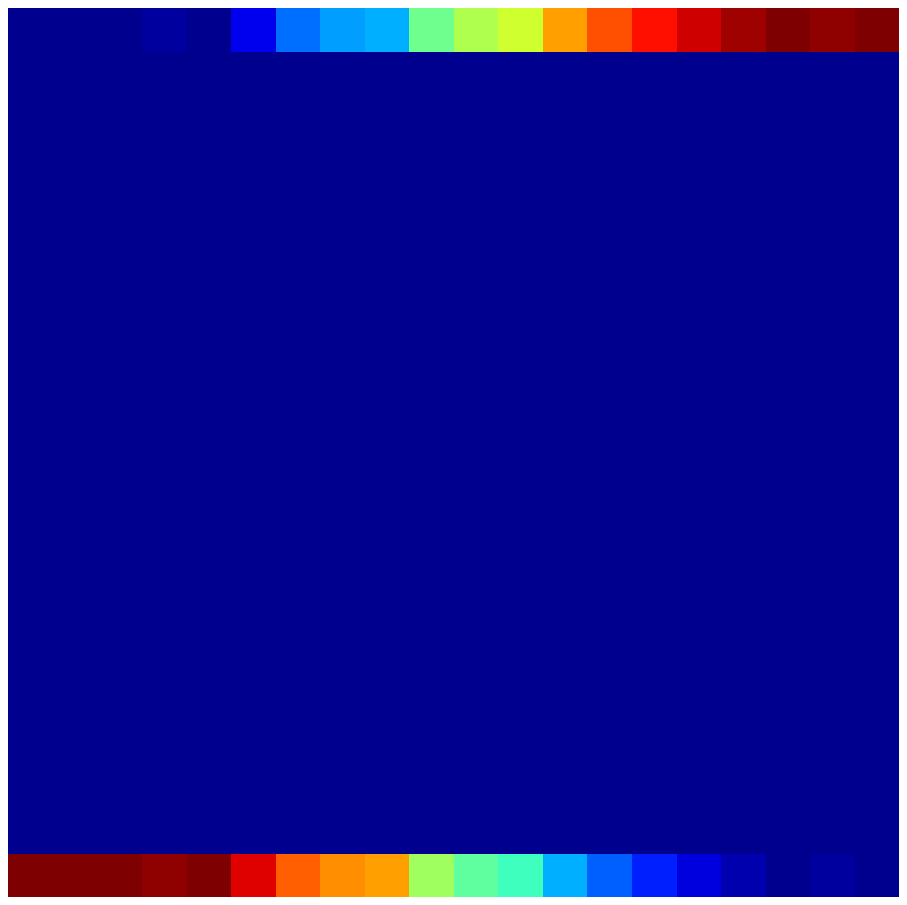}
\end{center}
\vglue -0.2cm
\caption{Density plot of probability $W_f$ to find a final red fraction $f_f$,
shown in $y$-axis, in dependence on
an initial red fraction $f_i$, shown in $x$-axis; data are shown inside
the unit square $0<f_i,f_f<1$. The values of $W_f$ are
defined as a relative number of realizations
found inside each of $20\times20$ cells,
which cover the whole unit square.
Here $N_r=10^3$ realizations of randomly distributed red
and blue colors are used to obtain
$W_f$ values (with convergence time up to $t=150$).
Here $a=0.2$ (left top panel),
$0.5$ (left bottom panel),
$0.65$ (right top panel),
$ 0.8$ (right bottom panel); $N=10^4$.
The probability $W_f$ is proportional to color
changing from zero (blue) to unity (brown).
\label{fig2}}
\end{figure}

For a comprehensive analyzes of the dependence of the final fraction
of red nodes $f_f$ on the initial state $f_i$, we consider below the evolution
of $f(t)$ for a large number of $N_r$ initial (random) distributions
of red nodes (Fig.~\ref{fig2}). We find that there is a certain
critical value $f_c$ such, that initial fractions $f_i$
of red nodes completely die out if $f_i<f_c$,
or become dominant for $f_i>1-f_c$.
For $a=0.2$ the value of $f_c$ is $f_c\approx 0.45$,
while for $a=0.65$ we have $f_c\approx 0.35$. In contrast to results obtained in
\cite{prof} we find that the system has no bistability
for $a<0.7$: the final state
is fixed for a concrete homogeneous initial distribution of opinions.
However, for a dominating tenacious society at $a>0.7$ there is
a small probability that a small initial fraction of red nodes
leads to a complete domination of red color
for values of $f_i>f_c$ (see Fig. \ref{fig2} left bottom panel).
For the case of $a=0.8$, we have $f_c\approx 0.3$.
Obviously, the results are symmetric with respect to a change of red and blue colors.

\begin{figure}[ht]
\begin{center}
\includegraphics[width=0.23\textwidth]{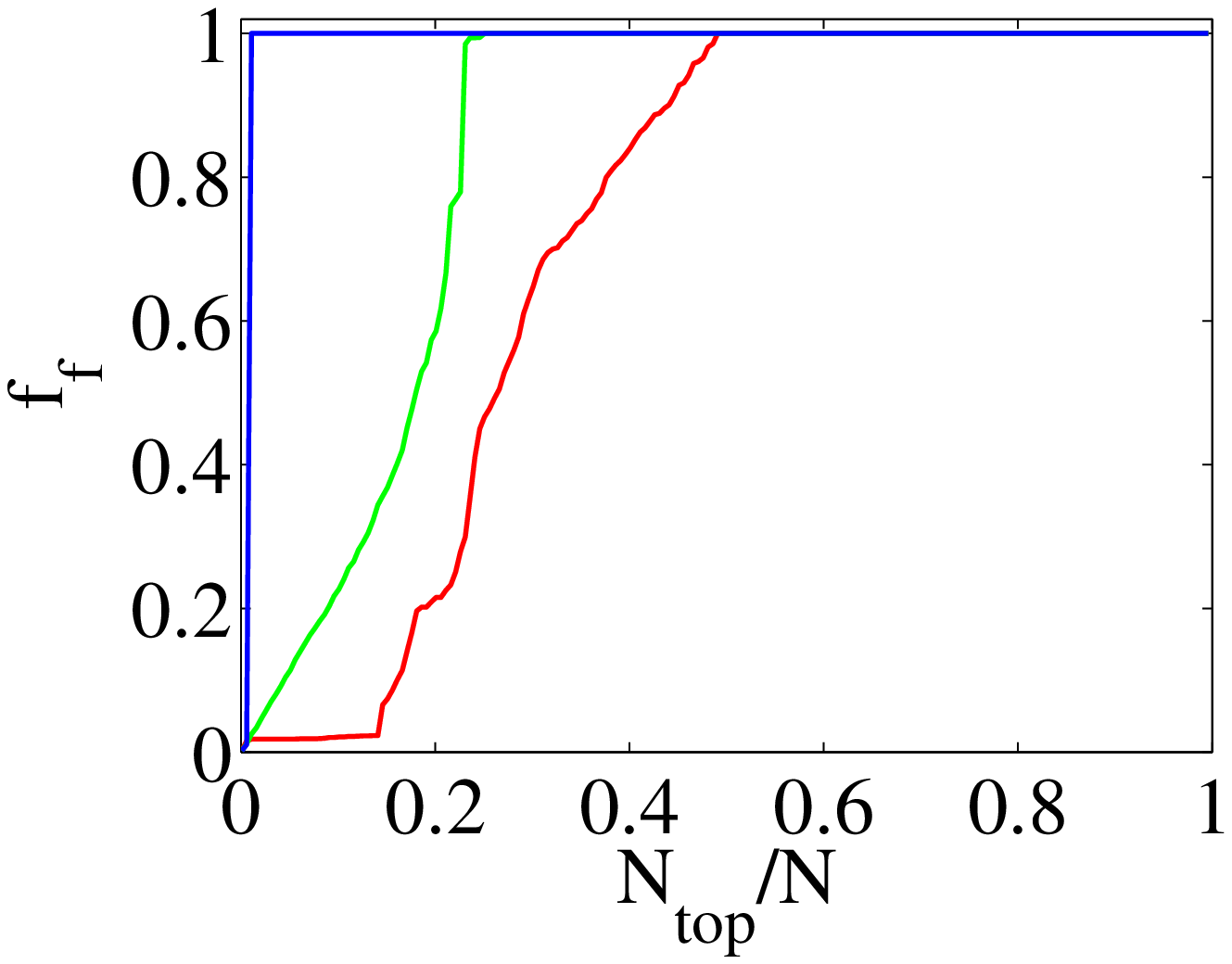}
\includegraphics[width=0.23\textwidth]{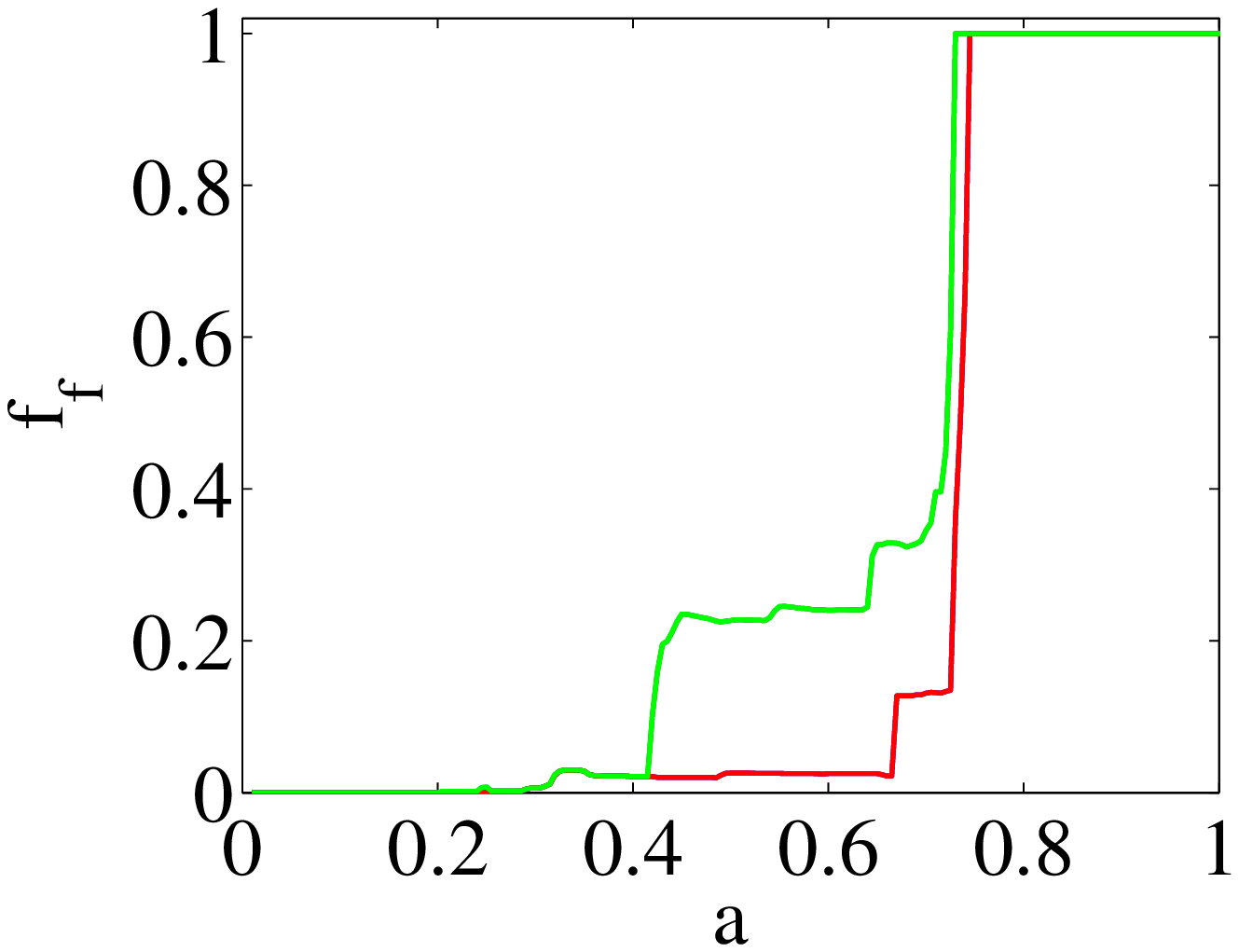}
\end{center}
\caption {Left panel: final fraction of red nodes $f_f$ versus $N_{top}/N$,
for  $a=0.4$ (red), $0.6$ (green), $0.8$ (blue).
Right panel: dependence of the final fraction of red nodes
$f_f$ on the parameter $a$, for initial state with different number
of the first $N_{top}$ nodes of the highest PageRank being red:
$N_{top}=100$ (red); 1000 (green). Here $N=10^4$. \label{fig3}}
\end{figure}

We also analyze how the final state depends on the number of
the elite members $N_{top}$ with the highest PageRank of
the same opinion (Fig.~\ref{fig3}).
We see that for any type of a society (any $a$) there exists
a value of $N_{top}^c$ such that the elite can convince the whole society,
if $N_{top}>N_{top}^c$. Note that the value of $N_{top}^c$ depends on the
tenacious parameter $a$. The larger the tenacious parameter is,
the smaller number of the elite members of
a same opinion can bring the system to unanimity.

\section{The generalized PROF-Sznajd model}
\label{sznajd}

In this section we consider the properties of the combination of PROF
and Sznajd models \cite{sznajd}. The Sznajd model features the idea of groups
of a society and thus incorporates a well-known principle
"United we stand, divided we fall". A thorough analyzes of the problem
on regular lattice networks can be found in Ref.~\cite{revsznajd}.
The present generalization (which results in the PROF-Sznajd model) is applicable to
scale-free and Ulam networks. We define the notion of group
of nodes at each discrete time step $\tau$ following Ref.~\cite{prof}:

\begin{enumerate}

\item we pick randomly a node $i$ in the network and consider
the state of the $N_g-1$ highest PageRank nodes pointing to it;

\item if the node $i$ and all other $N_g-1$ nodes have the same color
(same spin orientation), these $N_g$ nodes form a group, whose effective
PageRank value is the sum of all the member values $P_g=\sum_j^{N_g}P_j$.
If it is not the case, we leave the
nodes unchanged and perform the next time step;

\item consider all the nodes pointing to any member of the group and check
all these nodes $n$ directly linked to the group: if an individual node PageRank
value $P_n$ is less than the defined above $P_g$, the node joins the group
by taking the same color (polarization) as the group
nodes and increase $P_g$ by the value of $P_n$; if it is not the case,
a node is left unchanged.
\end{enumerate}

\begin{figure}[ht]
\begin{center}
\includegraphics[width=0.23\textwidth]{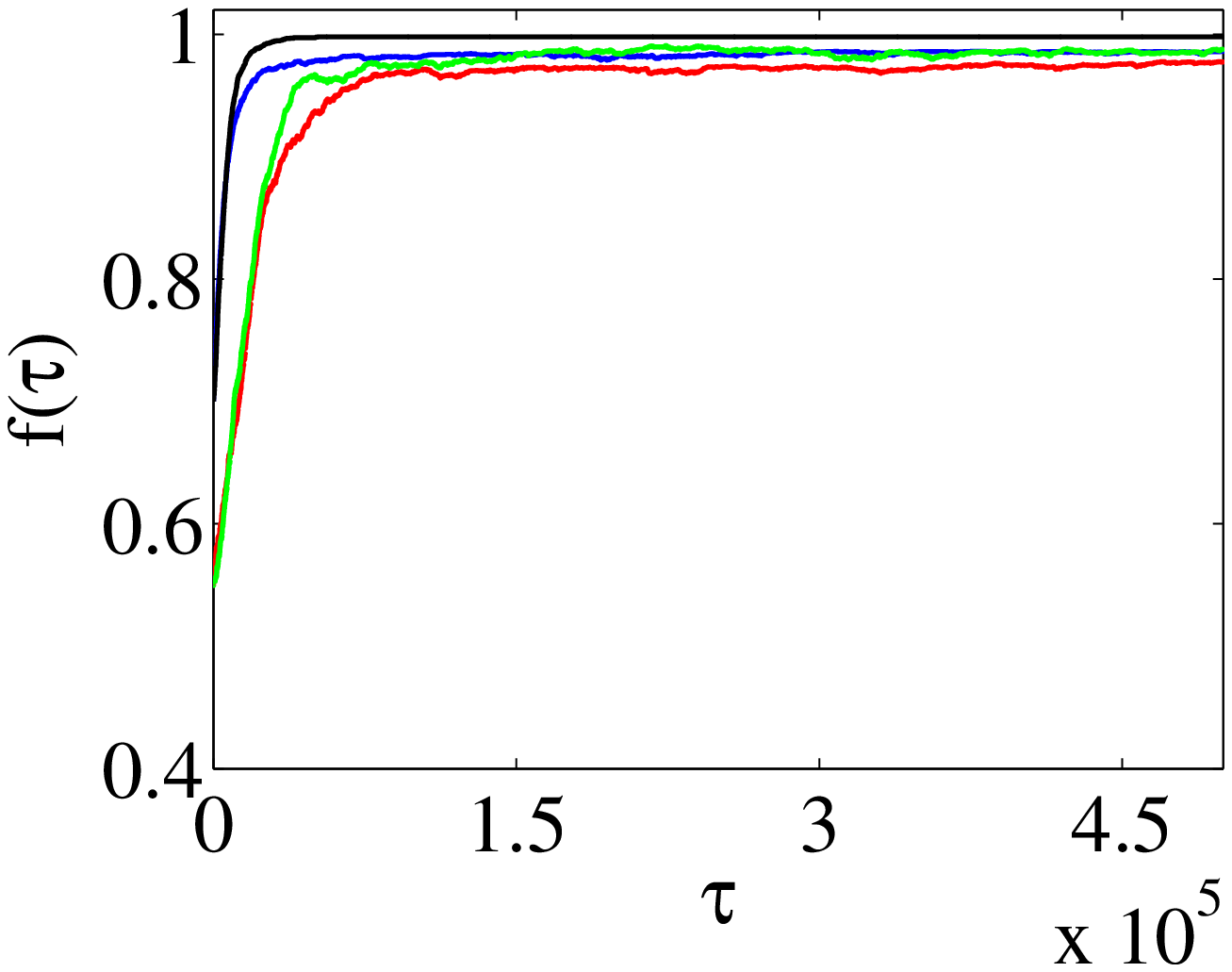}
\includegraphics[width=0.23\textwidth]{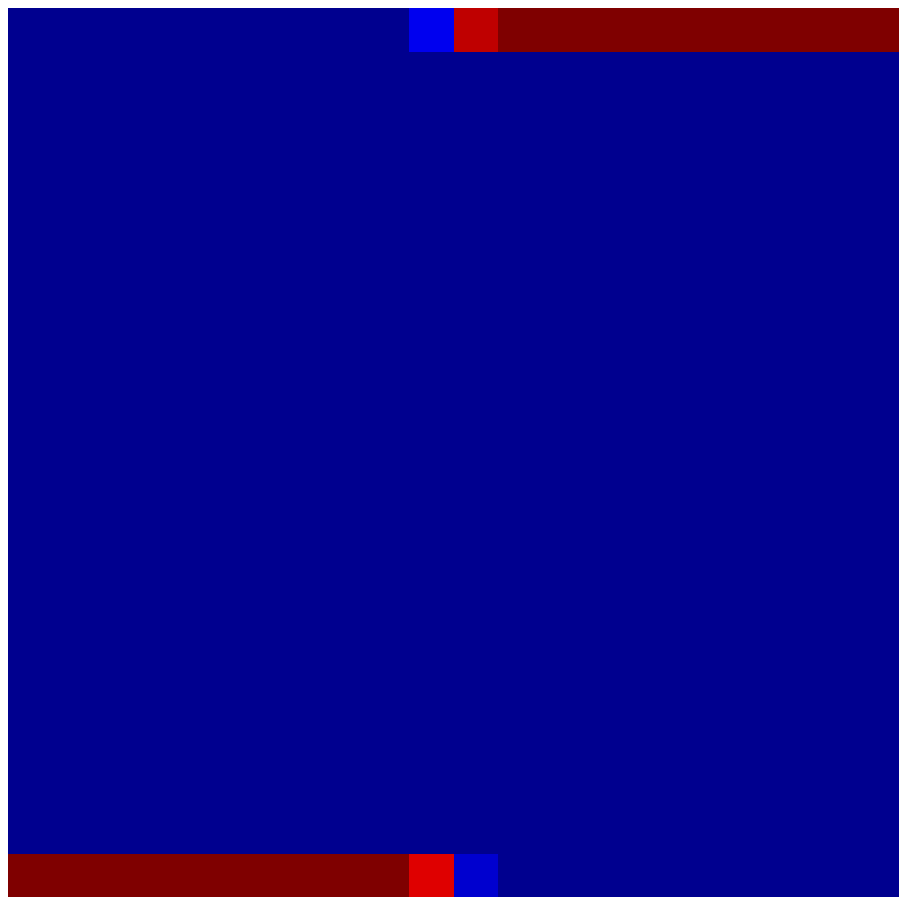}
\end{center}
\caption{Left panel: time evolution of the fraction of red nodes $f(\tau)$
of the PROF-Sznajd model, with different initial fractions of red nodes and
the group size $N_g$  (at one random realization each): $f_i=0.55$, $N_g=3$ (red);
$f_i=0.55$, $N_g=4$ (green); $f_i=0.7$, $N_g=3$ (blue); $f_i=0.7$, $N_g=4$ (black).
Right panel: the same as in Fig.~\ref{fig2},
but for the PROF-Sznajd model with group size $N_g=3$,
with convergence time up to $\tau=5\cdot 10^5$;
colors are as in Fig.~{\ref{fig2}}. Here $N=10^4$.
\label{fig4}}
\end{figure}

In Fig.~\ref{fig4} we present a typical behavior of the PROF-Sznajd model on
Ulam network generated by the intermittency map. Firstly, we find that the convergence
time is longer than that of the PROF model, which is the generic feature of
the Sznajd model. The system converges to its final state after a time $\tau_c$
of the order of $\tau_c\sim10 N$. Note that there are still some fluctuations
in the steady state regime, which were absent in the conventional PROF model.
Another observation concerns the group size $N_g$: we find that the size
of the group does not affect much the properties of the model:
there is a small decrease in the resistivity of minorities with
the group size increase (of around $2\%$ with a change from $N_g=3$ to $N_g=4$).
Furthermore, the network practically does not have nodes with more
than four incoming links, hence, we find that considering a group size
with $N_g>5$ loses its sense.

The right panel of Fig.~\ref{fig4} shows a density plot of probability $W_f$,
constructed in a similar to Fig.~\ref{fig2} way. We see, that the rate of
surviving of small fractions of (red) nodes is drastically small
(we address this result to the poor incoming link structure of the Ulam network).
The initial states are suppressed if $f_i\lesssim0.45$. But for $0.45<f_i<0.5$
($0.5<f_i<0.55$) there is a small probability of approximately $8\%$ that
the fraction will become dominant (be suppressed). Outside of this
small range of $f_i$ we don't find any regions of bistability:
the final state of the system is fixed.

For the PROF-Sznajd model we are additionally interested in the Ulam network,
generated by another dynamical map, the typical Chirikov map
with dissipation:

\begin{equation}
\left\{\begin{array}{l l }
y_{t+1}=\eta y_t+k \sin(x_t+\theta_t), \\ \nonumber
x_{t+1}=x_t+y_{t+1}.
\end{array} \right.\label{2}
\end{equation}

Here the dynamical variables $x$ and $y$ are taken at integer moments of time $t$.
Also $x$ has a meaning of phase variable and $y$ is a conjugated momentum or action.
For a detailed description of this dynamical system, see Ref.~\cite{ulamd}.
The map region is $0\leq x<2\pi$ and $-\pi\leq y<\pi$,
with $2\pi$-periodic boundary conditions. The phases $\theta_t=\theta_{t+T}$
are $T$ random phases periodically repeated along time $t$.
Here we consider  the  $T10$ case with $T=10$,
analyzed in Ref.~\cite{ulamd}. The values of parameters are set
to $\eta=0.99$, $k=0.22$. The list of 10 values of $\theta_t$
phases can be found in the Appendix of Ref.~\cite{ulamd}.
For the construction of the Ulam network we divide the phase space
to $n_x\times n_y$ cells ($n_x=n_y=100$). Afterwards, $N_c$ trajectories
are propagated from each given cell $j$ during $T$ map iterations
to obtain elements of the adjacency matrix $S_{ij}$
for transitions to cell $i$ (in the same manner as for the mapping (\ref{1})).
The total matrix size is $N=10^4$.

\begin{figure}[ht]
\begin{center}
\includegraphics[width=0.23\textwidth]{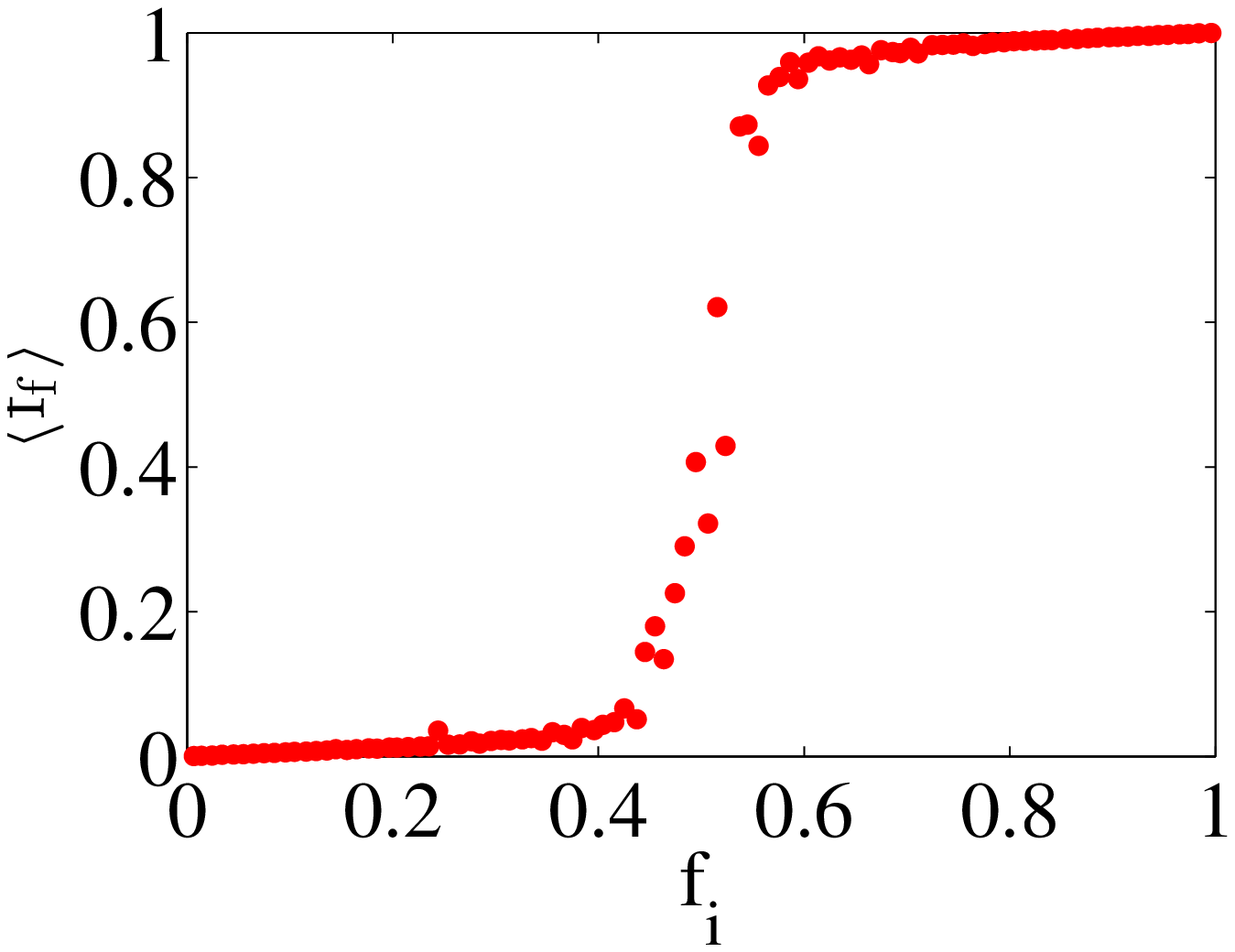}
\includegraphics[width=0.23\textwidth]{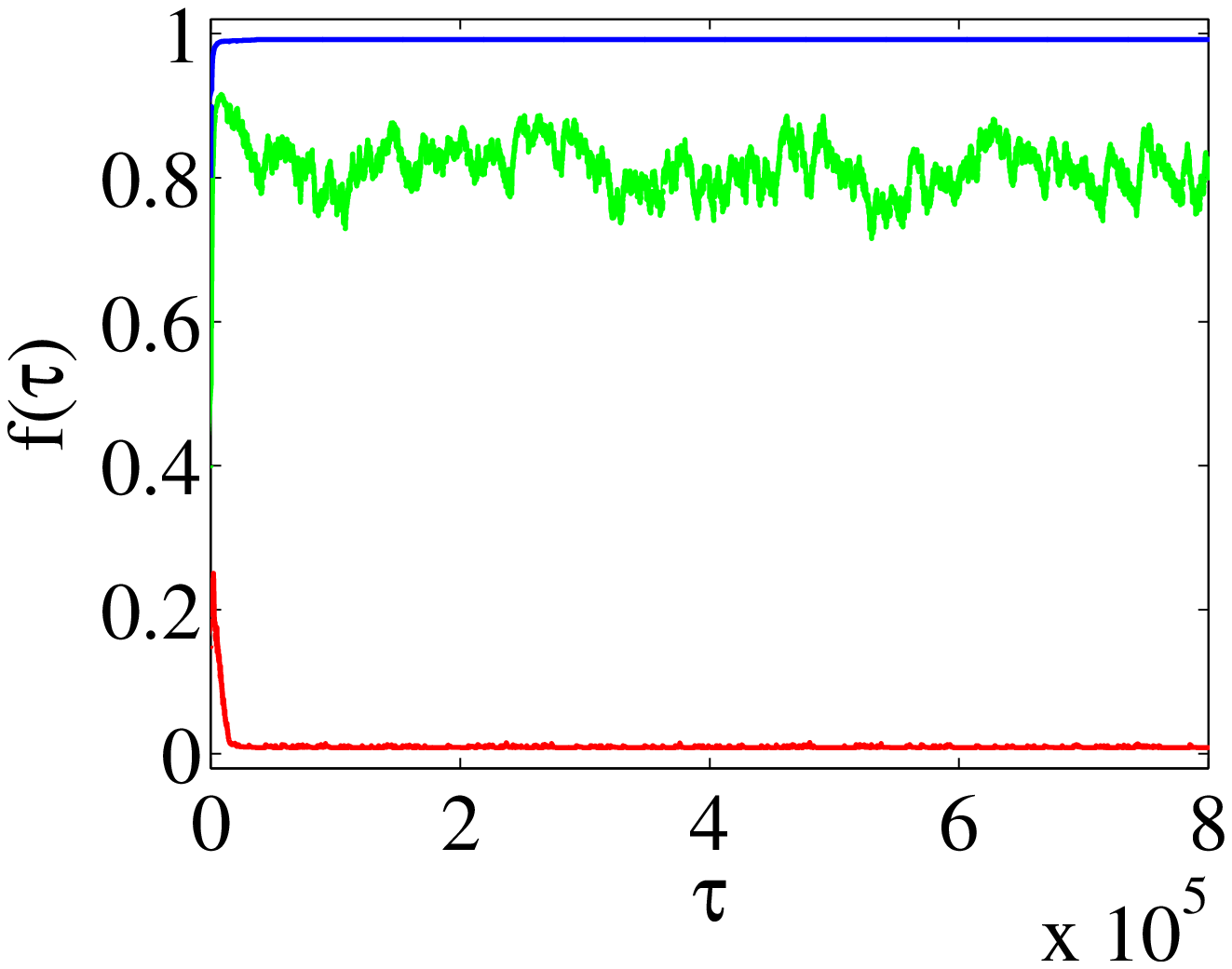}
\end{center}
\caption{Left panel: The average of the final fraction of red nodes
$\langle f_f \rangle$ versus the initial fraction $f_i$ for
the PROF-Sznajd model of T10 model of the typical Chirikov map
($\alpha=0.95$, $n_x=n_y=100$, $N=10^4$). Here, $N_r=10^3$ realizations
with a convergence time up to $\tau=3\cdot 10^5$ are used to
obtain the average $\langle f_f \rangle$ (the group size is $N_g=3$).
Right panel: time evolution of the fraction of red nodes $f(\tau)$ for
the same model, for the initial state with the first $N_{top}$
nodes of the highest PageRank being red: $N_{top}=1500$ (red);
$N_{top}=4000$ (green); $N_{top}=8000$ (blue).
\label{fig5}}
\end{figure}

For this network we find a higher strength of resistivity of minorities,
since it has a richer link structure. On Fig.~\ref{fig5} we plot the average
of the final fraction of red nodes $f_f$ versus the initial fraction $f_i$.
We see here that minor opinions die out if $f_i\lesssim 0.3$. The damping factor
of the Google matrix here is set to $\alpha=0.95$, which gives
a power law decay of the PageRank with a slope of $0.48$ (see Ref.~\cite{ulamd}).
We also looked at the $f_f$ versus $f_i$ behavior for other values of
the damping factor. As mentioned above, the Google matrix properties of
Ulam networks are sensitive to the values of $\alpha$. Nevertheless, our
calculations showed, that for $0.95<\alpha<1$, qualitative behavior of
the PROF-Sznajd model remains similar to that of Fig.~\ref{fig5}. On the other hand,
as pointed out in Ref.~\cite{prof}, the increase of the slope of
the power law decay of the PageRank should result in a bistable behavior
of the PROF and PROF-Sznajd models on social and university networks.
However, this argument does not hold true for Ulam networks: although
the slope of the PageRank increases with growth of $\alpha$
(e.g. for $\alpha=0.98$ we have $P_j\propto1/j^{0.7}$,
while for $\alpha=0.99$ we have $P_j\propto1/j^{0.9}$), bistability does not emerge.
Thus we conclude that a presence of bistability behavior is associated
not only with the slope of the PageRank decay, but also
with the intrinsic structure of the network itself.

For the PROF-Sznajd T10 model we find
that the elite of the society cannot convince any elector,
if its fraction is initially relatively small. In particularly,
the first $N_{top}$ nodes of the highest PageRank with the same opinion
are suppressed for  $N_{top}/N\lesssim0.2$.
For $N_{top}/N\gtrsim0.2$, the elite becomes capable to
influence the opinion of other electors, but the convergence process
as well as the final state starts exhibiting fluctuations of a
significant  amplitude.
These fluctuations become smaller for higher values of $N_{top}$
and almost disappear for $N_{top}/N\gtrsim0.7$
where the society comes to unanimity.

\begin{figure}[ht]
\begin{center}
\includegraphics[width=0.23\textwidth]{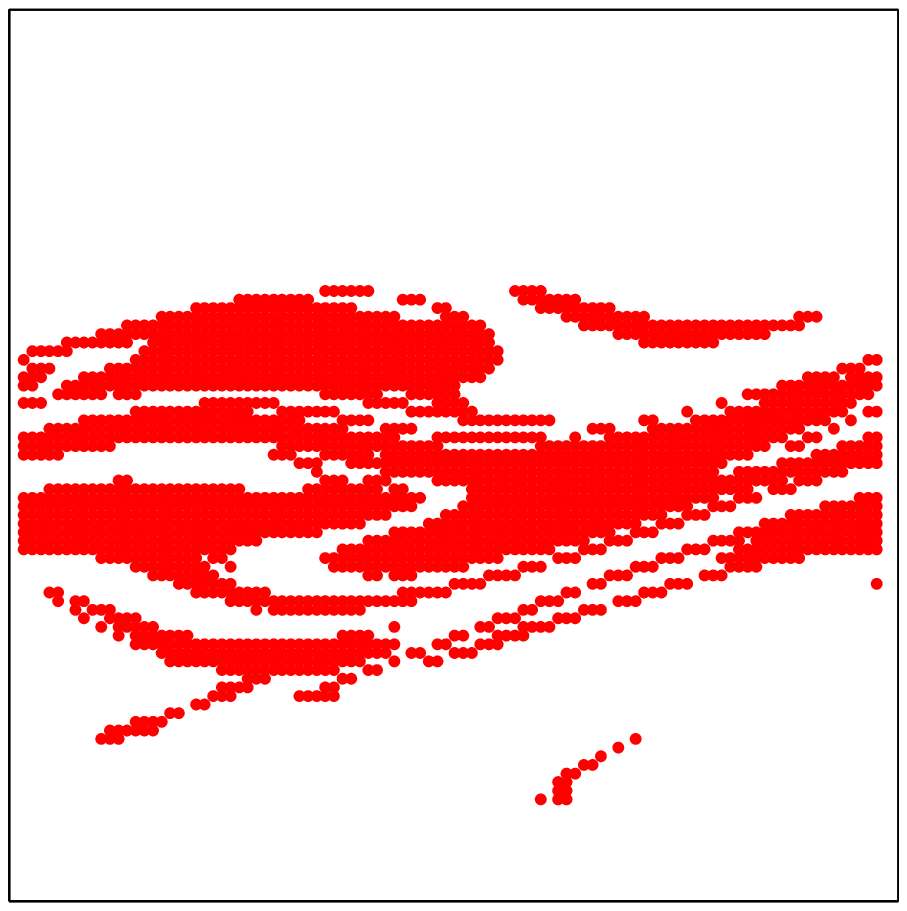}
\includegraphics[width=0.23\textwidth]{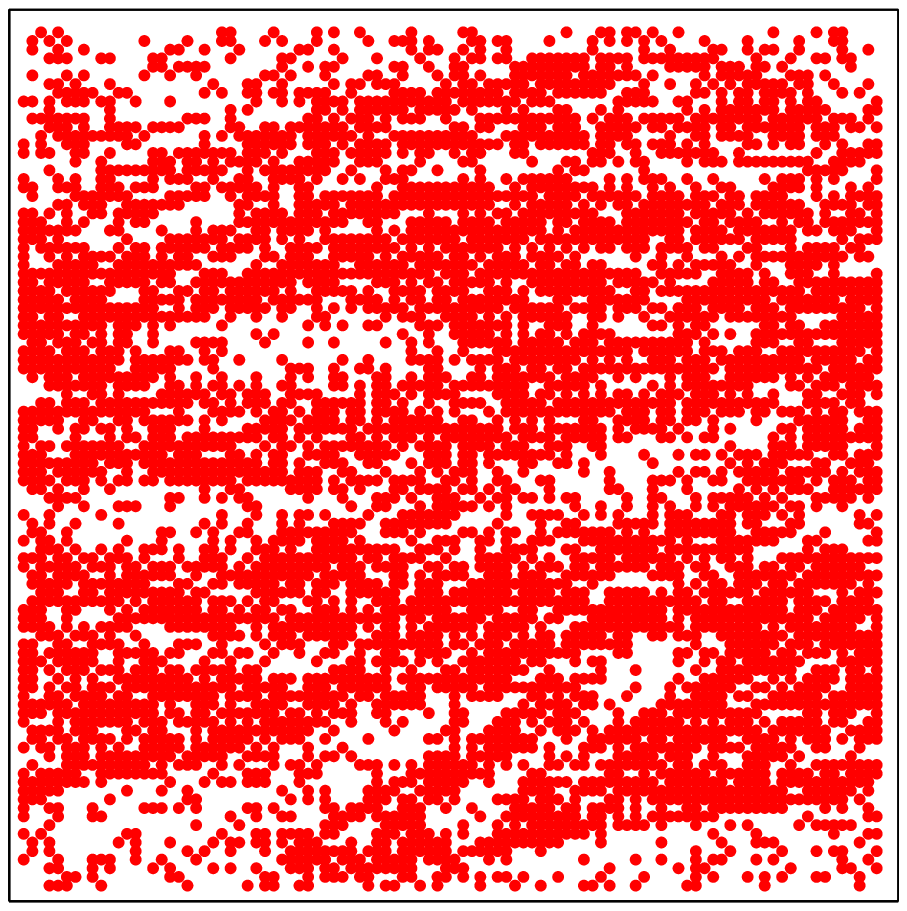}
\end{center}
\vglue -0.2cm
\caption{Coordinate distribution of red nodes in $(y, x)$ phase space of
the PROF-Sznajd $T10$ model ($\alpha=0.95$, $n_x=n_y=100$, $N=10^4$);
the phase plane is shown in $2\pi \times 2\pi$ square.
Left panel shows the initial state with $N_{top}=2200$ nodes of
the highest PageRank being red and $f_i=0.22$;
right panel corresponds to the final state
with $f_f=0.5758$.
\label{fig6}}
\end{figure}

Finally, we shortly describe the initial and final distributions
of red nodes in the coordinate space. It is of interest to consider the case
of initial state with $N_{top}$ red nodes with the highest PageRank,
since for random distributions the final and initial states
are homogeneously distributed over phase plane. Figure~\ref{fig6} shows
the initial and final distributions for $N_{top}=2200$.
We find that the top elite nodes first tend to convince other members of the elite
corresponding to the denser regions on the right panel of
Fig.~\ref{fig6} with high values of the PageRank probability.

\section{Discussion}
\label{concl}

In this work we analyzed the features of a recently proposed PageRank opinion formation
model on two examples of Ulam networks. The Ulam networks generated by
the discussed above one dimensional intermittency and
typical Chrikov maps exhibit some intrinsic properties similar to the WWW.
This fact makes the analyzes relevant to the opinion formation process in real societies.
We pointed out that the elite of a society does not have a considerable influence
on the decision making process of the electors for an equal mixture of conformist
and tenacious society. However, the influence of the elite becomes tangible
for a dominating tenacious society. In contrast to the university networks analyzed in
\cite{prof} we find practically no regions of
bistability behaviour for a random distribution of initial opinions.
Only  a dominating tenacious society shows some signs of bistability.

We also considered a generalization of the Sznajd model for Ulam networks
(PROF-Sznajd model). We found here that the system still practically does not
feature bistable regimes. On the basis of our studies we conclude
that the PageRank decay exponent does not influence
the bistability for the Ulam networks considered in this work.
We argue that the chaotic maps considered generate strong stretching
of small regions of phase space but do not generate
significant number of loop returns. We think that this feature
is different from university networks which are characterized by a
significant number of loops. We presume that this internal feature
of the Ulam networks is at the origin of significant difference
in opinion formation on these two types of scale-free networks.
The presented results can be useful for analysis of opinion formation
on other types of scale-free directed networks.

\section*{Acknowledgments}
We thank N.Ananikyan for useful discussions.
This work was supported by the France-Armenia collaboration grant CNRS/SCS
No. 24943 (IE-017) on "Classical and quantum chaos" and
EC FET Open project NADINE $N 288956$.
L.C. gratefully acknowledges the funding by the Conseil R\'{e}gional
de Bourgogne and FP7/2007-2013 grant No. 205025-IPERA.

\end{document}